\documentclass[10pt,a4paper,twocolumn,english,american,aps,prl,twocolumn,showpacs,floatfix,superscriptaddress,10pt]{revtex4-1}
\usepackage[T1]{fontenc}
\usepackage[latin9]{inputenc}
\usepackage{babel}
\usepackage{bm}
\usepackage{array}
\usepackage{float}
\usepackage{multirow}
\usepackage{amssymb}
\usepackage{graphicx}
\usepackage[normalem]{ulem}

\usepackage{xcolor}
\definecolor{grey}{rgb}{.6,.6,.6}

\PassOptionsToPackage{normalem}{ulem}
\usepackage[unicode=true,
 bookmarks=true,bookmarksnumbered=true,bookmarksopen=true,bookmarksopenlevel=2,
 breaklinks=false,pdfborder={0 0 0},backref=false,colorlinks=true]
 {hyperref}
\hypersetup{pdfauthor={Maxen Cosset-Ch\'eneau },pdfsubject={Spintronics}}
\usepackage{breakurl}


\begin{document}

\title{Electrical measurement of the Spin Hall Effect isotropy in a ferromagnet}

\author{M. Cosset-Ch\'eneau, M. Husien Fahmy, A. Kandazoglou, \\C. Grezes, A. Brenac, S. Teresi, P. Sgarro, P. Warin, A. Marty,\\
 V.T. Pham, J.-P. Attan\'e, L. Vila}
\affiliation{Universit\'{e} Grenoble Alpes, CEA, CNRS, INP-G, Spintec, F-38054 Grenoble, France}
\date{\today}
\selectlanguage{american}%
\begin{abstract}
The spin-dependent transport properties of paramagnetic metals are roughly invariant under rotation. By contrast, in ferromagnetic materials the magnetization breaks the rotational symmetry, and thus the spin Hall effect is expected to become anisotropic. Here, using a specific design of lateral spin valves, we measure electrically the spin Hall Effect anisotropy in NiCu and NiPd, both in their ferromagnetic and paramagnetic phases. We show that the appearance of the ferromagnetic order does not lead to a sizeable anisotropy of the spin charge interconversion in these materials.  

\end{abstract}
\maketitle

Spintronics is primarily based on the exchange interaction between the spins of conduction electrons and the local magnetization in ferromagnets (FMs). This interaction is fundamental in effects such as the Giant Magnetoresistance \cite{ValetPRB1993} and the Spin Transfer Torque \cite{StilesPRB2002}. More recently, the use of Spin Orbit (SO) interactions to manipulate spins in non-magnetic (NM) materials has triggered the birth of spinorbitronics, a new subfield of research and technology aimed at developing new spintronics devices based on SO effects, such as the SOT-MRAM \cite{ShaoIEEE2021} or the MESO devices \cite{ManipatruniNature2019}. \\
The main SO effect used in spinorbitronics is the Spin Hall Effect (SHE) \cite{SinovaRMP2015}, that allows the conversion of a spin current into a charge current. So far, the study of the SHE has been mostly focused on non-ferromagnetic systems, such as semiconductors \cite{KatoScience2004} and heavy metals \cite{SinovaRMP2015}.\\  
Recent experiments and theoretical calculations have however shown that beyond their usual role for spin current injection \cite{TserkovnyakPRB2002, LaczkowskiPRB2019} and detection \cite{ValetPRB1993, LaczkowskiPRB2019, LiuPRL2011, ChenPRB2013} through exchange coupling, FMs possess  an interesting potential for spin-charge interconversion \cite{OmoriPRB2019, GladiiPRB2019, TaniguchiPRA2015, KimataNature2019, HibinoNC2021}.
The presence of the magnetization in these materials raises a fundamental question, as it breaks the rotational symmetry \cite{DavidsonPLA2020}. The SHE in FMs is therefore expected to be anisotropic, i.e., it should depend on the relative orientation between the magnetization and the applied charge current \cite{AminPRB2019}. 

The rare observations of this anisotropy have been made in 3d elements below the Curie temperature, but show different results for similar materials, with a moderate anisotropy in Py \cite{DasPRB2017}, a large one in CoFe \cite{WimmerAPL2019}, and an isotropic behaviour in Co \cite{TianPRB2016}. This could result from differences in-between material properties, but also from the fact that a precise measurement of this anisotropy remains challenging. The main techniques used to measure the SHE in heavy metals are spin pumping \cite{AndoAPL2011}, local \cite{PhamNL2016} and non-local \cite{ValenzuelaNature2006} transport in nanostructures, Spin Hall Magnetoresistance \cite{KimPRL2016}, Spin Seebeck effect \cite{UchidaNature2008} and Spin-Orbit torque techniques such as ST-FMR \cite{LiuPRL2011} or second harmonic \cite{HayashiPRB2014}. Observing the SHE anisotropy in FMs using these methods is however not straightforward, as it requires controlling independently the magnetization of two different FMs in direct contact \cite{VarottoPRL2020, OuNL2019, GibbonsPRA2018, TianPRB2016, MiaoPRL2013, Zhu2020NPGAM}. 
Techniques involving a charge current flowing into the FM-SHE material are made complicated by contamination from spurious effects such as the Planar Hall effect \cite{MizuoAPL2021} and Anomalous Hall Effect (AHE) \cite{GroenPRA2021}, while the methods based on non-local magnon transport that allowed for the observation of an anisotropic SHE in FMs \cite{DasPRB2017, WimmerAPL2019} involve the direct contact between the magnetic insulator and the studied FM material, which could render complex the interpretation of these results \cite{CramerPRB2019}.\\

Here, we propose a new design of lateral spin valves, which provides a way to measure the anisotropy of the SHE in FMs. We use it to measure this anisotropy in NiCu and NiPd, which have been found to exhibit spin Hall angles (i.e., the charge to spin conversion rate) similar to those of the best heavy metals \cite{VarottoPRL2020, KellerPRB2019,WeiNatCom2012}. The low Curie temperature of these materials allow performing the measurements either in the paramagnetic or ferromagnetic phase. In both cases, the measured inverse spin Hall effect signal is found to be independent from the magnetization direction, the observed variations being smaller than our measurement precision of 10\%. These results demonstrate that in these materials the breaking of symmetry due to the ferromagnetic order does not lead to a significant anisotropy of the SHE.

\begin{figure}[h]
    \centering
    \includegraphics[scale=0.55]{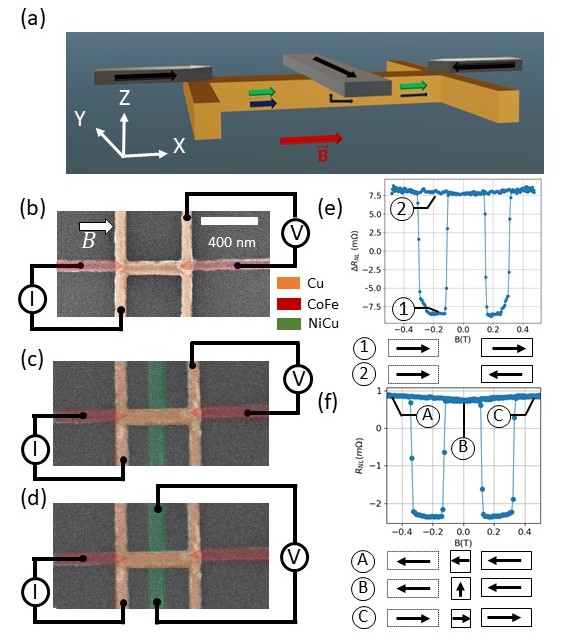}
    \caption{(a) Device geometry, with the FM parts in gray, and the Cu in orange. The black arrows in the FM parts represent the magnetization directions. The dark blue and green arrows within the channel represent the spin current flow direction and its polarization, respectively.  The LSV has been represented with the FMs on top for the sake of clarity, but in the actual device they lie below the Cu channel to ensure well-defined micromagnetic configurations within the electrodes and in the absorber. (b) Colored SEM image of a reference LSV without absorber, in the NL measurement configuration. (c) Colored SEM image of a LSV with absorber, in the NL measurement configuration. (d) Colored SEM image of a LSV in the ISHE measurement configuration. (e) NL signal at 12 K for a reference LSV and (f) with a 100 nm wide NiCu absorber.}
    \label{fig:fig1}
\end{figure}

%%%%%%%%%%%%%%%%%%%%%%%%
The measurements are based on the injection of a charge current from the injecting CoFe electrode towards one branch of the copper channel (cf. Fig. 1b). This creates a pure spin current flowing towards the detecting electrode (Fig. 1a). The device can be studied using two different measurement configurations. The non-local (NL) spin signal configuration consists in detecting the voltage drop in between the detecting electrode and the channel. By reversing the magnetization of the injecting electrode, using an external magnetic field along the easy magnetization axis, it is possible to change the polarization direction of the spin current. This leads to two different non-local voltages, high (1) and low (2), at the detecting electrode (Fig. 1e).
By comparing this signal with that of a device with absorber (Fig. 1c and 1f), it allows measuring the spin current flowing towards the absorber \cite{ZahndPRB2018}. It is then possible to use the ISHE measurement configuration (Fig. 1d), in which the spin current is partly absorbed by the SHE absorber. The polarization of the absorbed spin current is along $X$, whereas the absorbed spin current direction flows along $Z$ axis. It thus generates in the ferromagnetic SHE absorber an electric field, along $Y$, and consequently an ISHE signal at the extremities of the absorber.
%%%%%%%%%%%%%%%%%%%%%%%%%%%

When applying small fields along $X$, the magnetization of the ferromagnetic SHE absorber rotates, while the magnetization of the electrodes remain along $X$. It is thus possible to control the angle in between the polarisation direction of the absorbed spin current, and the magnetization of the absorber \cite{SM}. Then, if one applies large enough magnetic fields along $X$, the magnetization of the injecting electrode reverses. The polarization of the injected spin current is then reversed. This causes a sign change of the ISHE signal/spin signal, allowing to measure the ISHE/spin signal amplitude.\\

The main advantage of this detection scheme with respect to methods based on LSVs \cite{OmoriPRB2019, NiimiRPP2015} is the possibility to control independently the magnetization direction of the electrodes and of the absorber, moreover using very small fields, in order to avoid field-induced contributions such as the Hall or Hanle effects \cite{SanchezAPL2013, ZahndAPL2018}.

We used CoFe injecting and detecting electrodes, and Cu for the copper channel, in order to take advantage of its long spin diffusion length and small interface resistance with CoFe \cite{ZahndNanotech2016, LaczkowskiPRB2019}. A control sample has also been realised using a platinum absorber \cite{SM}, as Pt is a well known material  \cite{VilaPRL2007}. Then, the measurements of the SHE anisotropy were performed in devices with absorbers made of NiCu and NiPd. As these materials are paramagnetic at room temperature and ferromagnetic at low temperatures, it is thus possible to measure if the appearance of ferromagnetism leads to a SHE anisotropy.\\
The devices were patterned on PMMA, by conventional e-beam on a $\mathrm{SiO_2}$ substrate. The Cu, CoFe and Py wires were deposited by physical vapor deposition, while the NiCu was deposited by sputtering from a $\mathrm{Ni_{60}Cu_{40}}$ target followed by a lift-off, while $\mathrm{Ni_{16}Pd_{84}}$ was codeposited from a Ni and a Pd target. The CoFe electrodes were deposited in a first step, followed by the ISHE wire, and finally by the copper channel in a third step. We cleaned the interfaces \textit{in situ} by Ion Beam Etching, prior to the deposition of the copper channel.\\

\begin{figure}[h]
    \centering
    \includegraphics[scale=0.37]{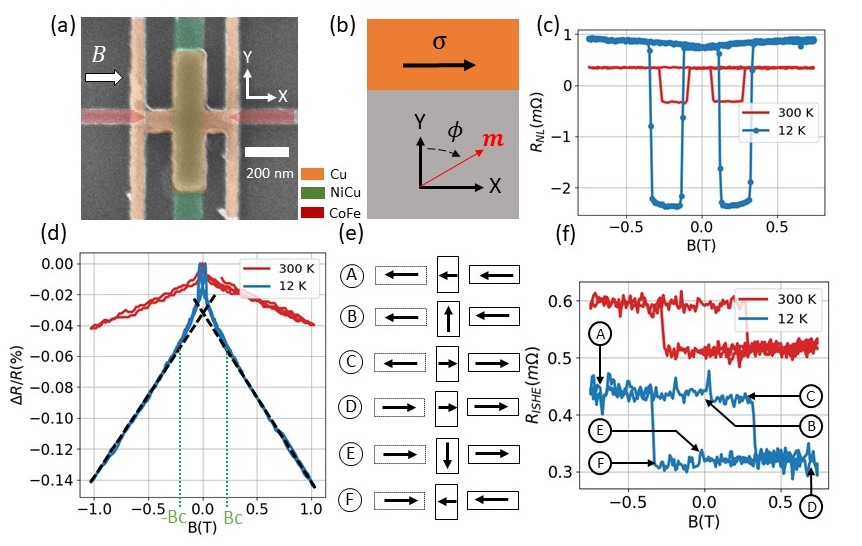}
    \caption{(a) colored SEM image of a LSV with a NiCu absorber. The central part of the device is additionaly covered with a Cu thin film, to avoid spurious magnetoresistive contributions to the signal \cite{SM}. (b) Relative orientation between the spin current polarization $\sigma$ and the magnetization of the absorber at low ($\phi=0$) and high ($\phi=\pi/2$) fields. (c) NL signal at room temperature (red) and at 12 K (blue) using a 100 nm wide NiCu absorber. (d) two probe magnetoresistance of the 100 nm wide NiCu absorber measured at 12 K (blue) and at room temperature (red), for a magnetic field transverse to its easy magnetization axis. (f) ISHE measurement in a NiCu absorber, made simultaneously to the NL measurement shown in (c), and (e) scheme of the corresponding magnetic configurations.}
    \label{fig:fig2}
\end{figure}

%%%%%%%%%%%%%%%%%%%%%%%%%%%%%%
The magnetoresistance measurements have been performed using lock-in techniques ($I=400 \, \mu$A, $f=123$ Hz). The non-local signal measured in a reference LSV without absorber (Fig. 1b) is larger (Fig. 1e) than the one detected in a device with a NiCu absorber (Fig. 1d and 1f), with a signal amplitude going from 15 m$\Omega$ to 3.3 m$\Omega$. This NL signal difference allows measuring the spin conductance of the absorbing SHE wire, and thus evaluating the amount of spin current absorbed by the ferromagnetic ISHE material \cite{SM}. %%%%%%%%%%%%%%%%%%%%%%%%%%%%%%%%%%

In the device presented in Fig. 2a, we performed both non-local measurements (cf. Fig. 1c) and ISHE measurements (Fig. 1d), with an ISHE signal amplitude of 120 $\mu\Omega$. Contrarily to what can be obtained using classical LSVs with FM materials \cite{OmoriPRB2019}, the signals can here be measured at low fields, allowing the control of the 
angle $\phi$ in between the absorber magnetization and the spin current polarization (cf. Fig. 2b). At low fields, the magnetization of the absorber is along $Y$, and thus transverse to the spin current polarization, which is along $X$. 
At high fields, the magnetization of the absorber is along X, and thus collinear to the spin current polarization.

Let us firstly focus on the non-local measurements. Using the measurement configuration of Fig. 1c, a clear non-local signals can be obtained for devices with NiCu absorbing wire, both in the ferromagnetic and the paramagnetic phase (cf. fig. 2c). 

Adding the absorbing NiCu wire to the LSV leads to a strong decrease of the non-local signal (cf. Figs. 1e vs. 1f). The presence of the absorbing wire between the injecting and detecting CoFe electrode indeed leads to a partial absorption of the spin current flowing into the channel. The absorbed spin current is denoted $j_z^x$, as it flows in the $Z$ direction and has a spin polarization along $X$. Its expression is given by the relation \cite{ChenPRB2013}:
\begin{equation}
    j_z^x=G_{\mathrm{eff}} (\boldsymbol{\mathrm{m}}) \mu^x
\end{equation}
with $\hat{m}$ the magnetization direction of the absorber and $\mu^x$ the spin accumulation along the $x$ direction at the interface between the Cu channel and the absorber. $G_{\mathrm{eff}}$ is treated as an effective parameter taking into account the spin conductance $G_\mathrm{s}=1/\rho l_{\mathrm{sf}}$ with $\rho$ and $l_{\mathrm{sf}}$ the resistivity and spin diffusion length of the absorbing material \cite{Cosset-CheneauPRL2021} (for $\boldsymbol{\mathrm{m}}=\hat{x}$), or the spin mixing conductance $G_{\uparrow \downarrow}$ ($\boldsymbol{\mathrm{m}}=\hat{y}$ or $\hat{z}$) \cite{BrataasPR2006, Cosset-CheneauPRL2021}, as well as the interface resistance $R_{int}$. 
 
We extracted from the NL signals $G_{\mathrm{eff}}$ and thus $j_z^x$ at the NiCu/Cu interface, using FEM simulations \cite{LaczkowskiPRB2019} (see \cite{SM} for computation details and extracted parameters). As seen in Fig. 2c, the non-local signal at 12 K, and thus the absorbed spin current, is found to be relatively independent on the relative angle between the spin polarization and the absorber magnetization \cite{SM}. This absence of modulation, which contrasts with the results obtained in ref. \cite{Cosset-CheneauPRL2021} using Py, Co and CoFe as ferromagnetic absorbers, is possibly due to the relatively high interface resistance ($7 \, \mathrm{f}\Omega \cdot \mathrm{m}^2$) between Cu and NiCu \cite{SM} when compared to the minimal Sharvin interfacial resistance at the copper 3d-FM interface ($0.8 \, \mathrm{f}\Omega \cdot \mathrm{m}^2$) \cite{Cosset-CheneauPRL2021}. The modulation being smaller than the experimental error bar, the values of the spin conductance and of the effective spin mixing conductance can be considered to be equal.

Now that the absorbed spin current is known, let us focus on the ISHE. We firstly measured the dependence of the magnetization direction of the NiCu SHE absorber as a function of the field applied along X, by performing two probe magnetoresistance measurements of the NiCu wire (Fig. 2d). The magnon magnetoresistance contribution \cite{MihaiPRB2008, NguyenPRL2011} dominates the signal in both the paramagnetic and ferromagnetic phases. In the ferromagnetic phase, the magnetoresistance also displays at low fields a deviation from the linear behavior, corresponding to the expected anisotropic magnetoresistance. This allows measuring the saturation field of the NiCu wire ($B_c=0.2$ T) along the $X$ axis ($\phi=\pi/2$).

We then measured the ISHE signal using the configuration of Fig. 1d, at both 12 K and 300 K. The Curie temperature of $\mathrm{Ni_{60} Cu_{40}}$ has been found to be around $220$ K for this composition \cite{VarottoPRL2020}. Using the non-local and ISHE measurements, we extracted an effective spin Hall angle ($2.3$\%) consistent with the one of ref. \cite{VarottoPRL2020} measured using spin pumping, which confirms the validity of our measurement method \cite{SM}. The ISHE signal, shown in Fig. 2f, is a square loop in both the paramagnetic and ferromagnetic phases. The observed square loops reflect the change of sign of the absorbed spin current polarization when reversing the injector magnetization. The symmetry and shape of the signal correspond to what can be expected from the ISHE symmetries, and from the switching fields measured in the non-local configuration. The parallel and transverse ISHE signals can then be obtained as $\Delta R_{\mathrm{ISHE}}^{\perp}=R^B-R^E$ and $\Delta R_{\mathrm{ISHE}}^{||}=R^A-R^D$, $R^i$ being the electric potential drop in the magnetic state $i$ reported in Fig. 2e. The key finding is that at 12 K, the ISHE signal is constant in the [-0.2 T; 0.2 T] field range, where the magnetization of the absorber rotates from the $X$ to the $Y$ direction, so that $\Delta R_{\mathrm{ISHE}}^{\perp}=\Delta R_{\mathrm{ISHE}}^{||}$. This means that the ISHE is independent on the direction of the absorber magnetization.
This isotropy of the ISHE in the ferromagnetic phase is also illustrated by the fact that very similar loops in shape are observed in the paramagnetic phase, where one does not expect any anisotropy, and in control devices using Pt as absorber \cite{SM}. 

\begin{figure}[h]
    \centering
    \includegraphics[scale=0.45]{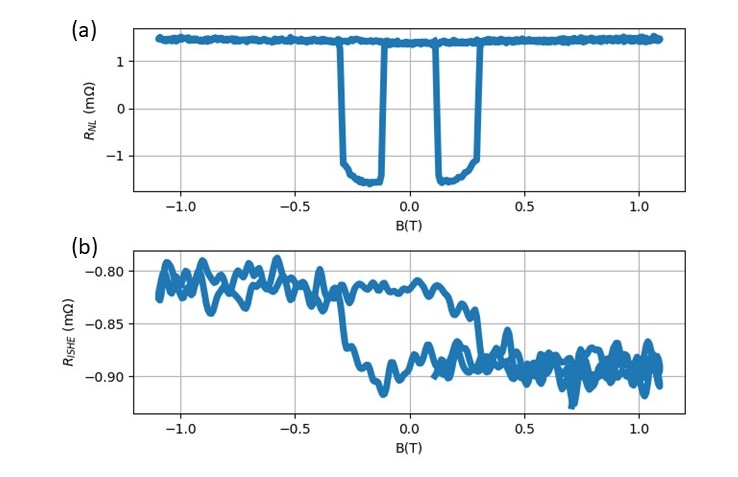}
    \caption{((a) Non-local and (b) ISHE (d) signals measured during the same magnetic field sweep, at 150 K with a 100 nm NiPd absorber.}
    \label{fig:fig3}
\end{figure}

We also performed similar measurements using a 15 nm thick $\mathrm{Ni_{16}Pd_{84}}$ ferromagnetic absorber, in which the origin of the ISHE is extrinsic \cite{WeiNatCom2012} while it is intrinsic in NiCu \cite{VarottoPRL2020}. This alloy is ferromagnetic below 230 K. It has an out of plane magnetization at low temperature \cite{SM}, but in-plane above 150 K (cf. Fig. 3c). In order to avoid the strain induced magnetic inhomogeneity associated with an out of plane magnetization in this material \cite{ArhamPRB2009}, we measured the ISHE and non-local signal at 150 K when NiPd is still magnetic, and its magnetization is in plane. Similarly to the case of NiCu, the ISHE signal is found to be isotropic with respect to the magnetization direction.

%%%% discussion
The ISHE thus appears to be isotropic both for NiCu and NiPd in their ferromagnetic phase. This result is very surprising, as there is a symmetry breaking induced by the appearance of the magnetization at the paramagnetic/ferromagnetic transition. Theoretical studies have indeed predicted a sizeable modulation of the interconversion when rotating the magnetization \cite{AminPRB2019, QuPRB2020}. 

Let us now look at the links between this isotropy of the signal and the spin transport parameters.
The absorbed spin current Jc is converted into a charge current $J_\mathrm{s}$ according to $J_\mathrm{c}=\theta J_\mathrm{s}$. Importantly, $\theta$ is not the sole figure of merit of the ISHE. The conversion indeed occurs only over a limited material thickness, roughly equal to the spin relaxation length, after which the spin current, which is not conserved, vanishes. The produced signal is thus actually proportional to the product of the spin Hall angle by the spin relaxation length. If the incoming spins are collinear to the magnetization, this product is $\theta_{//} l_{\mathrm{sf}}$, where $\theta_{//}$ is the spin Hall angle in the collinear situation. If the incoming spins are transverse to the magnetization, this product becomes $\theta_\perp$ times $\lambda_\perp$, with $\lambda_\perp$ the transverse spin relaxation length. $\lambda_\perp$ represents the ballistic relaxation by phase averaging of the spin current with a polarization transverse to the magnetization.
The isotropy of the ISHE signal suggests that in our experiments:
\begin{equation}
\theta_{//}l_{\mathrm{sf}}\sim\theta_{\perp}\lambda_\perp    
\end{equation}
In principle, it is possible to measure separately $l_{\mathrm{sf}}$ and $\lambda_\perp$, as they affect the spin absorption \cite{Cosset-CheneauPRL2021}.
$l_{\mathrm{sf}}$ has been measured in NiCu \cite{VarottoPRL2020} and is equal to 2.4 nm at room temperature. The measurement of $\lambda_\perp$ is more complex. The presence of a large interface resistance prevents here the observation of the spin absorption anisotropy in our devices. It also has the effect of keeping the amount of spin current contributing to the ISHE constant. Extracting $\lambda_\perp$ would require thickness dependence measurement \cite{Zhu2020NPGAM} with all the spurious effects linked film thickness reduction \cite{GonzalezPRB2021}.

An estimation of $\lambda_\perp$ can be obtained by assuming that it is inversely proportional to the exchange field \cite{PetitjeanPRL2012}, and thus to the Curie temperature \cite{TurekPM2006}.
By taking as reference material Py, which has a Tc of 800 K \cite{ZhangAPL2019} and an exchange constant of $J_{\mathrm{Py}} \sim 100$ meV with $\lambda_\perp^{\mathrm{Py}} \sim 1$ nm \cite{PetitjeanPRL2012,Cosset-CheneauPRL2021}, $\lambda_\perp$ in NiCu can be estimated to be $\lambda_\perp^{\mathrm{NiCu}} \sim 5$ nm with $J_{\mathrm{NiCu}} \sim 20$ meV. The transverse relaxation length due to ballistic effects in NiCu is therefore larger than its spin diffusion length. This means that the transverse spin current relaxes mainly over the spin diffusion length through diffusive processes and that $\lambda_\perp^{\mathrm{NiCu}}=l_{\mathrm{sf}}$. 
$\theta_{//}$ has been estimated to be 2.3 \% \cite{SM}, which would lead according to our experimental result of eq. (2) to the same value for $\theta_\perp$. The case of NiPd is more difficult to analyse since we do not have values of the spin diffusion length in this material. However, the combination of the large SO interaction evidenced by the out of plane magnetic anisotropy at low temperature, and small exchange interaction evidenced by the low Curie temperature makes us expect a similar relation between the transverse spin relaxation and the spin diffusion lengths.\\

In light of these results, it appears that the isotropy of the absorption is to be expected in low-$T_\mathrm{c}$ FMs with strong spin-orbit coupling since the low exchange interaction drives the transverse spin relaxation length to be equal to the spin diffusion length. Hence, from the eq. (2) we can conclude that the spin Hall angle in NiCu and NiPd is isotropic. The picture in high-$T_\mathrm{c}$ FMs is however more contrasted. Indeed, it has been established that a strong anisotropy of the absorption exists in these materials \cite{Cosset-CheneauPRL2021}.  The isotropy of the ISHE signal has been observed in CoFe \cite{WimmerAPL2019} and Py \cite{DasPRB2017}, where the relation given by eq. (2) does not seem to hold \cite{Zhu2020NPGAM}, while it does in Co \cite{MiaoPRL2013}. According to theoretical studies \cite{AminPRB2019}, the details of the SOC effect on the band structure of these materials would provide the explanation for these behaviors, and we believe that the device described in this paper can provide an important tool for a systematic investigation of these phenomena. \\
To conclude, we proposed a device allowing the measurement of the anisotropy of the spin-charge interconversion in ferromagnetic materials. Although the paramagnetic/ferromagnetic transition is associated to symmetry breaking, the ISHE signal in NiCu and NiPd remains surprisingly independent of the magnetization direction in the ferromagnetic phase. 
In terms of spin transport properties, this implies that the product of the SHA by the spin relaxation length is the same in the colinear and transverse configuration. In the studied low $\mathrm{T_c}$ materials, this can be understood by the long transverse relaxation length which causes the spin current to relax during scattering on SO impurities and therefore to be equal to the spin diffusion length. The isotropy of the spin Hall angle is however not explained by these considerations and will require a more systematic theoretical and experimental work.

We acknowledge the financial support by ANR French National Research Agency OISO (ANR-17-CE24-0026). This project also received funding from the EUropen Union's Horizon 2020 research and innovation program under the Marie Slodowska-Curie Grant Agreement No. 955671.

The data associated to the figures can be found at the following link https://doi.org/10.57745/Z3BG2U.

%\end{multicols}
\end{document}